\documentclass[aps,preprint,showkeys,nofootinbib,showpacs]{revtex4}
\usepackage{amsmath,graphicx}
\newcommand{\bneqn}{\begin{equation}}
\newcommand{\edeqn}{\end{equation}}
\newcommand{\bneqna}{\begin{eqnarray}}
\newcommand{\edeqna}{\end{eqnarray}}
\newcommand{\bneqnas}{\begin{eqnarray*}}
\newcommand{\edeqnas}{\end{eqnarray*}}
\def\la{\langle}\def\ra{\rangle}
\def\be{\begin{eqnarray}}\def\ba{\begin{eqnarray}}
\def\ee{\end{eqnarray}}\def\ea{\end{eqnarray}}
\def\ben{\begin{enumerate}}\def\bitem{\begin{itemize}}
\def\een{\end{enumerate}}\def\eitem{\end{itemize}}
\def\no{\nonumber\\}

\begin{document}
\setcounter{page}{1}
\title[]{Isospin Matter in AdS/QCD}
\author{Kyung-il \surname{Kim}}
\affiliation{Institute of Physics and Applied Physics,
Yonsei~University, Seoul 120-749, Korea}
\email{hellmare@yonsei.ac.kr}
\author{Youngman \surname{Kim}}
\affiliation{Asia Pacific Center for Theoretical Physics and
Department of Physics, Pohang University of Science and Technology,
Pohang, Gyeongbuk 790-784, Korea,} \affiliation{School of Physics,
Korea Institute for
  Advanced Study, Seoul 130-012, Korea}\email{ykim@apctp.org},
\author{Su Houng \surname{Lee}}
\affiliation{Institute of Physics and Applied Physics, Yonsei
University, Seoul 120-749, Korea}
 \email{suhoung@yonsei.ac.kr}
\date[]{}


\begin{abstract}

We study strange and isospin asymmetric matter in a bottom-up
AdS/QCD model. We first consider isospin matter, which has served as
a good testing ground for nonperturbative QCD. We calculate the
isospin chemical potential dependence of hadronic observables such
as the masses and the decay constants of the pseudo-scalar, vector,
and axial-vector mesons. We discuss a possibility of the charged
pion condensation in the matter within the  bottom-up AdS/QCD model.
Then, we study the properties of the hadronic observables in strange
matter. We calculate the deconfinement temperature in strange and
isospin asymmetric matter. One of the interesting results of our
study is that the critical temperature at a fixed baryon number
density increases when the strangeness chemical potential is
introduced. This suggests that if matter undergoes a first-order
transition to strange matter, the critical temperature shows a
sudden jump at the transition  point.

\end{abstract}


\maketitle

\section{Introduction}
Dense matter is one of the most challenging problems of modern
physics. Understanding the properties of such matter is important
for the physics of relativistic heavy-ion collisions and dense
stellar objects such as neutron stars. Recent developments based on
Gauge/Gravity correspondence about Anti de Sitter space (AdS) and
Conformal field theory (CFT)~\cite{adscft} have opened a new way to
study such matter in the framework of a holographic model of Quantum
chromodynamics (QCD), or AdS/QCD~\cite{TD,EKSS,PR}, see
Ref.~\cite{Erdmenger:2007cm} for a review. Also, there have been
many examples of such studies on dense nuclear
matter~\cite{denseMatter} and isospin matter~\cite{Andrei, KKL,
APSZ}.

Isospin matter, where $\mu_I$ (isospin chemical potential) is finite
and $\mu_B$ (baryon chemical potential) is zero,  was proposed by
Son and Stephanov~\cite{SSiso} as a useful setting to improve our
understanding of cold dense QCD. Although a QCD system with finite
$\mu_I$ and zero $\mu_B$ hardly exists in  nature, it has many
interesting and useful features. The standard lattice QCD technique
is applicable to isospin matter, unlike to the QCD at finite baryon
chemical potential. Furthermore,  we can analytically study the
system at very low $\mu_I$ by using chiral perturbation theory  and
at very high $\mu_I$ by using perturbative QCD. Therefore, it seems
quite natural that one adopts the system to test a theoretical tool
for its suitability to tackle the nonperturbative physics of QCD.

At the same time, strange matter has been of great interest ever
since the original discussions that it could be
stable~\cite{Witten84} and form strangelets\cite{Farhi84}. Although
no proof for its existence has been found yet, heavy ion collisions
at the Large Hadron Collider (LHC) could lead to its discovery as
strange quarks will be amply produced if a quark-gluon plasma is
formed~\cite{Rafelski82}. Matter with strangeness chemical
potentials is also of interest in relation to neutron stars and the
phases of QCD at high density.

In this work, we first consider isospin matter, which has been
well-studied by using lattice QCD and effective theories of QCD, to
test our tool~\cite{EKSS, PR}. Some part of the present work is
briefly reported in Ref.~\cite{KKL}. We calculate the pseudo-scalar,
vector and axial-vector meson masses in isospin matter and discuss
the validity region of the model. We also evaluate the
$\mu_I$-dependence of the meson decay constants. Then, we study the
strangeness chemical potential dependence of the hadronic
observables and  investigate a possible onset of strange matter or
hyperonization in the AdS/QCD model. Finally, we study the effect of
the isospin and the strangeness number densities on the
deconfinement transition temperature.

\section{AdS/QCD Model with Various Chemical Potentials}

The action of the model given in Ref.\cite{PR, EKSS} is \ba
 S_5&=&\int d^4 x \int dz \mathcal{L}_5 \nonumber \\
 &=& \int d^4 x \int dz \sqrt{g}M_5 Tr [
-\frac{1}{4}(L_{MN}L^{MN}+R_{MN}R^{MN})\nonumber
\\ && \quad\quad\quad\quad+\frac{1}{2}|D_M \Phi |^2 -\frac{1}{2}M_{\Phi}^2 |\Phi|^2],\label{S5}
\ea where $M_5 = N_c/12 \pi^2$ and the field strength tensor is
defined by $ L_{MN} =
\partial_M L_N -\partial_N L_M + i[L_M,L_N]$  with gauge fields
$L_M=L^a_M t^a$ (and similar definition on $R_M$). Here, ${\rm
Tr}(t^a t^b)=\delta^{ab}$. Also, the covariant derivative has the
form $D_M \Phi =
\partial_M \Phi + i L_M \Phi - i \Phi R_M$. The
background is given by
\begin{equation}
ds^2 = a^2 (z) \left( \eta_{\mu\nu} dx^{\mu} dx ^{\nu} -dz^2
\right),
\end{equation}
where $a(z) = 1/z$ and $\eta_{\mu\nu} = diag(1,-1,-1,-1)$. Note that
we don't consider the back-reaction due to the chiral condensate to
the AdS background. In the model~\cite{EKSS, PR}, which is sometimes
called the hard-wall model, the AdS$_5$ space is compactified such
that $z_0\leq z \leq z_m$. Here, $z_0$ is the UV cutoff and $z_m$ is
for the IR cutoff. In Ref.\cite{EKSS, PR}, the IR cutoff, $z_m$, is
fixed by the rho-meson mass. The vector- and axial-vector mesons are
defined by
\begin{eqnarray}
&&V_M = \frac{1}{\sqrt{2}}(L_M+R_M), \nonumber \\
&&A_M = \frac{1}{\sqrt{2}}(L_M-R_M)\, .
\end{eqnarray}
The bulk scalar field is defined by  $\Phi = Se^{iP/v(z)}$, where
 $S$ and $P$ correspond to a real scalar and a pseudoscalar, respectively.
Here, the vacuum expectation value of S, $v(z)$, depends on the
quark mass matrix $M$ and the chiral condensate $\Sigma$, \ba v(z)=
M z +\Sigma z^3\, . \ea In Ref.\cite{EKSS}, it is assumed that
$\Sigma=\sigma {\bf E} $ and $M=m_q{\bf E}$, where ${\bf E}$ is the
two-by-two identity matrix for $N_f=2$. For $N_f=3$~\cite{RP06,
SW06, WX07, KS07}, $M$ and $\Sigma$ are assumed to be $M={\rm diag}
(\hat m, \hat m, m_s)$ and $\Sigma={\rm diag}
(\sigma,\sigma,\sigma)$ or $\Sigma={\rm diag}
(\sigma,\sigma,\sigma_s)$, where $\hat m=(m_u+m_d)/2$. In
Ref.\cite{SW06}, with the choice of $m_u=m_d\neq m_s$ and $\la\bar
uu \ra=\la\bar dd\ra\neq\la\bar ss\ra$, the strange quark sector of
the hard-wall model is extensively studied, and the result shows
good agreement with the experiment and the lattice QCD results,
except for the strange pseudoscalar mass.
 In Ref.\cite{SW06, WX07}, a deformed AdS background due to the
meson action is investigated with $N_f=3$.

 Now, we discuss how to introduce various chemical potentials into our model.
For this, we first consider the QCD side, where the chemical
potentials are introduced as follows: \ba
 {\cal L}_{\rm QCD}^\mu&=&\mu_u u^\dagger u +\mu_d d^\dagger
d+\mu_s s^\dagger s\, \no &=&(\mu_q+\mu_I) u^\dagger u
+(\mu_q-\mu_I) d^\dagger d+\mu_s s^\dagger s\, , \label{QCDm} \ea
where $\mu_q=\frac{1}{2}(\mu_u+\mu_d)$ and
$\mu_I\equiv\frac{1}{2}(\mu_u-\mu_d)$. According to an AdS/CFT
dictionary, a chemical potential in 4D QCD is encoded in the
boundary value of the time component of the 5D bulk U(1) gauge
field. To this end, we generalize the 5D gauge symmetry of the
model~\cite{EKSS, PR} from SU(3)$_{\rm L}\times$SU(3)$_{\rm R}$ to
U(3)$_{\rm L}\times$U(3)$_{\rm R}$~\cite{DH}. As shown in
Ref.\cite{DH}, the time component of the bulk U(1) vector field has
the following profile: \ba V_0=c_1 +c_2z^2\, . \ea Then, $c_1$ is a
chemical potential $\mu$, and $c_2$ is a conjugate number density
$\sim \rho$. If $c_1$ is the quark chemical potential, then $c_2$
must be the quark number density, $c_2=12\pi^2 n_q/N_c$~\cite{DH}.

We remark here that although the AdS/CFT dictionary states that the
integration constants of the vector profile must be interpreted as a
chemical potential and its conjugate density, it does not fix the
values or the relation of the chemical potential and the
corresponding density, which is a limitation of the present
approach. In this work, we consider the baryon (or quark) chemical
potential, the isospin chemical potential, and the strangeness
chemical potential. To include such chemical potentials in the
model, we rewrite the time component of the 5D bulk U(1) field as
\ba V_0 &=&\hat{V_0}\hat{t}+V^a_0t^a\no &=&\hat{V_0}\hat{t}+V^8_0t^8
+ V_0^3 t^3 +\dots \no
 &=& 2\tilde{V_0} \tilde{t}+\sqrt{2}\bar{V_0}\bar{t}+ V_0^3 t^3
+\dots\, ,\label{V0p} \ea where $t^a$ are the generators of SU(3),
$\hat t$ is the generator of the U(1) subalgebra of U(3), $\hat
t=diag(1/\sqrt{3},1/\sqrt{3},1/\sqrt{3})$, and \ba \quad \tilde{t}=
\frac{1}{\sqrt{2}}\left(
\begin{tabular}{ccc}1&&\\&1&\\&&0\end{tabular}\right),\quad \bar{t}= \left(
\begin{tabular}{ccc}0&&\\&0&\\&&1\end{tabular}\right),\quad t^3= \frac{1}{\sqrt{2}}\left(
\begin{tabular}{ccc}1&&\\&-1&\\&&0\end{tabular}\right)\, .
\ea {}From the form of redefined generators, we see that
$\tilde{V_0}\leftrightarrow (\tilde\mu_q,~ \rho_q)$,
$V_0^3\leftrightarrow (\tilde\mu_I,~\rho_I)$ and
$\bar{V_0}\leftrightarrow (\tilde\mu_s,~\rho_s)$. More explicitly,
we write\protect\footnote{We use $\tilde\mu$ instead of $\mu$
because our chemical potential may be different from the one in Eq.
(\ref{QCDm}) by an overall normalization constant.  } \ba {\tilde
V}_0=\tilde\mu_q+c_qz^2\,,~~{\bar V}_0=\tilde\mu_s+c_sz^2\, ,~~
V_0^3=\tilde\mu_I+c_Iz^2\, . \ea

Among the three chemical potentials, the most interesting one might
be the baryon (quark) chemical potential. To look at this, we put
$\tilde V_0$ into  the action in Eq. (\ref{S5}). However, $\tilde
V_0$ does not couple to the bulk fields in the action, simply
because the action contains only mesons~\cite{KT}. An exceptional
case is to consider the Chern-Simons couplings at finite baryon
density~\cite{DH}. Therefore, we consider the isospin matter to test
our model in the medium because the matter has been well-studied by
using various approaches~\cite{SSiso, NJL, Lattice, Other}.

\subsection{Isospin Matter}
To study isospin matter, we turn on only $V_0^3$ and $\tilde\mu_I$
and turn off the others in Eq. (\ref{V0p}). Then, the profile of the
isospin-triplet bulk vector field is given by \ba V_0^3t^3=
(\tilde\mu_I +c_I z^2 ) t^3\, . \ea Since we cannot determine the
relation between $\tilde\mu_I$ and $c_I$ within our framework, we
treat both of them as external parameters. For simplicity, we may
borrow the relation from QCD. For example, in Ref.\cite{SSiso}, it
is given by \ba &&c_I=0~: {\rm normal~ phase ~with~ no~ pion~
condensation }\, \no &&c_I=f_\pi^2\tilde\mu_I
(1-m_\pi^4/f_\pi^4)~:{\rm pion ~condensed~ phase  }\, .\nonumber \ea
Now, we calculate the pion masses at small $|\tilde\mu_I|$ and study
if pion condensation occurs as we increases $|\tilde\mu_I|$.
 For a small isospin chemical potential, we have
\ba
V_0^3t^3= \tilde\mu_I t^3\, .\label{V0}
\ea
Although  we are not able to describe the physics in the pion condensed phase with Eq.~(\ref{V0}),
we should be able to observe the onset of condensation, if it occurs.
 Note that if
$\mu_I$ is negative, pion condensation means $\pi^-$ condensation.
{}From the action in Eq. (\ref{S5}), we obtain the equation of motion of the pion field in isospin matter as
\begin{eqnarray}
\mathcal{D} \left( {m_\pm}^2 + (2 v^2 a^2 \mathcal{D} ) \pm 2\tilde\mu_I
m_\pm \right) \pi^{\pm} =0, \label{PionEoM} \\
 \mathcal{D} \left(
m_0^2 + (2 v^2 a^2 \mathcal{D} ) \right) \pi^0 =0\ ,
\end{eqnarray}
where
\begin{eqnarray}
 \mathcal{D} &\equiv& 1 - \partial_5 \left(
\frac{1}{2a^3 v^2 }
\partial_5 a \right)\, .\nonumber
\end{eqnarray}
Here $\pi^\pm$ are linear combinations of $\pi^1$ and $\pi^2$, and
$\pi^0$ is for the neutral pion, $\pi^3$. Note that the mass of the
neutral pion is independent of  $\tilde\mu_I$, as it should be,
because a neutral pion has a zero third component of the isospin. We
can immediately read off the $\tilde\mu_I$ dependence of the charged
pion masses from Eq.~(\ref{PionEoM}):
\begin{eqnarray}
m_{\pm} = \mp \tilde\mu_I + \sqrt{\tilde\mu_I^2 + m_0^2}\, ,
\end{eqnarray}
We find that the mass of  $\pi^+$ increases with $|\tilde\mu_I|$
while that of $\pi^-$ decreases as we raise $|\tilde\mu_I|$. At
small $\tilde\mu_I$, we have $m_{\pm} \approx \mp \tilde\mu_I +
m_0$, which is consistent with the observation made in
Ref.\cite{SSiso}.

Before we move on to the pion decay constant, we present some
remarks here on the pion condensation. In Ref.\cite{SSiso}, the
critical isospin chemical potential $\tilde\mu_I^c$ for the
condensation is determined by the condition that the mass of $\pi^-$
be zero. In the chiral limit, $m_0=0$, we always have the pion
condensation because $m_-$ is zero for any value of $\tilde\mu_I$,
which is consistent with the observation made in a top-down
approach~\cite{APSZ}. With nonzero $m_0$, however, the mass of
$\pi^-$ obtained in the present work is always non-zero, unless
$\tilde\mu_I$ goes to infinity. This means that in the present
set-up, we have no pion condensation, which is quite different from
what was observed in Ref.\cite{SSiso}, in the Nambu-Jona-Lasino(NJL)
model~\cite{NJL}, in Lattice~\cite{Lattice} and in other model
calculations~\cite{Other}. This may be understood as follows: At
small isospin chemical potential, we may ignore the back-reaction
due to $V_0^3$, or the isospin chemical potential, so the AdS metric
in Eq.(\ref{S5}) is good for describing the isospin matter. As we
increase the value of the isospin chemical potential, the effect of
the back-reaction of the isospin chemical potential on the metric is
no longer negligible, so our expectation is that if we consider the
back-reaction from the flavor sector (chemical potential or density)
and work with
 a deformed AdS$_5$, as for instance in Ref.\cite{Deformed},
then we are able to observe pion condensation in AdS/QCD. For pion
condensation in dense matter other than isospin matter, we refer to
Ref.\cite{pionC}.

\begin{figure}[!ht]
\begin{center}
{\includegraphics[angle=0, width=0.7\textwidth]{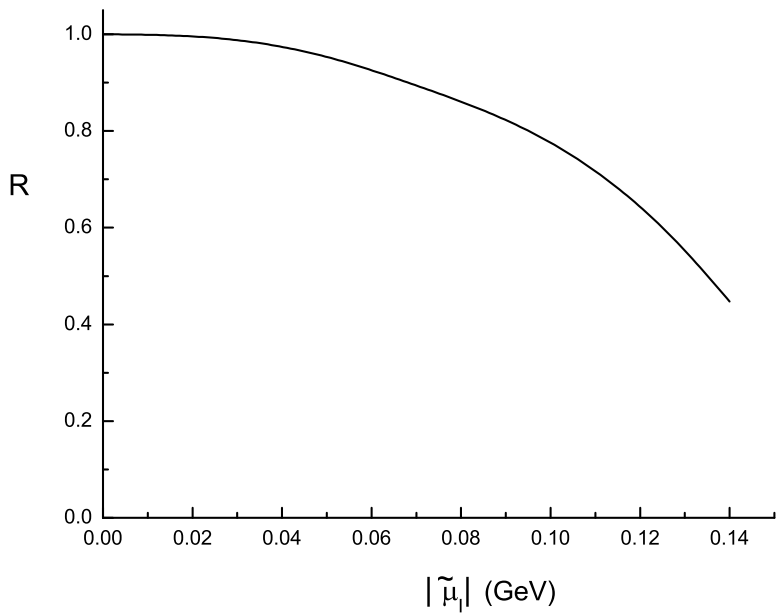} } \caption{
\label{fpis} \small  The spatial component of the pion decay
constant in isospin matter. Here, $R\equiv f_\pi^s (\tilde\mu_I
)/f_\pi(0)$. }
\end{center}
\end{figure}

Now, we calculate the $\tilde\mu_I$-dependence of the pion decay
constant. In matter, due to the lack of Lorentz invariance, we may
define the pion decay constant as~\cite{GYt}
\begin{eqnarray}
(f^{t,s}_{\pi})^2 = -\frac{1}{g^2}\frac{\partial_z A_{0,i}}{z}|_{z=\epsilon}\, ,\label{fpi}
\end{eqnarray}
where $\epsilon\rightarrow 0$ and $A_{0,i}$ is the solution of the
axial-vector equation of motion at zero 4D energy-momentum,
\begin{eqnarray}
[  a^{-1} \partial_5 a \partial_5 - 2v^2 a^2  +\tilde\mu_I^2]A_i =0\, .
\end{eqnarray}
Here, $A_i$ is the space component of the charged axial-vector
($A^1,~A^2$), and the equation of motion of the charge-neutral
axial-vector field does not depend on $\tilde\mu_I$. The
$\tilde\mu_I$-dependent $f_\pi^s$ is plotted in Fig.~\ref{fpis}.

Finally, we consider the vector and the axial-vector mesons,
especially the $\rho$ and the $a_1$ mesons that are the lowest
Kaluza-Klein (KK) states of the bulk vector field and the
axial-vector field respectively. The KK decomposition of the bulk
vector field is defined by  $V_\mu(x,z)=\frac{1}{\sqrt{M_5L}}\sum_n
f_n^V(z) V_\mu^{(n)} (x)$.

After integration by parts in the action, we obtain quadratic terms
for the vector field:
\begin{eqnarray}
\mathcal{L}_{\mu_I VV}&=&\frac{M_5 a }{2} \{ V^{\mp}_i [
\partial^2 - a^{-1} \partial_5 a \partial_5 \mp 2\tilde\mu_I i \partial_0
 - \tilde\mu_I^2 ] V^{\pm}_i  \nonumber \\ && + V^0_i[
\partial^2 - a^{-1} \partial_5 a \partial_5] V^0_i \}\, ,\label{EoMV}
\end{eqnarray}
where $V_0^3T^3= \tilde\mu_I T^3$ is used. Again, the charge-neutral
vector field is independent of $\mu_I$. {}From the quadratic action,
we find
\begin{eqnarray}
m_{\rho^{\pm}} =  m_{\rho^0}\mp\tilde\mu_I \, .
\end{eqnarray}
As expected, the negatively-charged $\rho$-meson mass decreases with
increasing $|\tilde\mu_I|$. Here we observe an interesting
possibility, namely, the vector meson condensation, because the mass
of $\rho^-$ is zero at $|\tilde \mu_I|=m_{\rho_0}$.( See
Ref.\cite{Vc} for a review on vector meson condensation.) As
discussed previously, to discuss the physics of condensation in the
present set-up, we may have to consider the back-reaction from the
isospin chemical potential to the background, so any definite
statement on the possibility of vector condensation has to be
postponed until we successfully obtain the back-reacted (deformed)
background. We refer to Ref.\cite{APSZ} for a discussion on vector
condensation in the chiral limit in a top-down model. We also
evaluate the $\rho$-meson decay constant in the isospin matter and
find it to be independent of $\tilde\mu_I$. This is obvious from Eq.
(\ref{EoMV}). The corrections due to finite $\tilde\mu_I$ to the
equation of motion for the vector field, $f_n(z)$, are
$z$-independent; consequently, the KK wave-function for the vector
field is blind to the isospin chemical potential. For the
axial-vector meson, $a_1$, we obtain the same results for the
$\rho$-meson: \ba m_{a_1^{\pm}} = m_{{a_1}^0}\mp\tilde\mu_I \, ,
\end{eqnarray}
and the $a_1$ decay constant is independent of $\tilde\mu_I$. To
complete the analysis, we need to calculate the
isospin-chemical-potential dependence of scalar mesons. The scalar
sector of the present model is, however, very sensitive to the
details of added scalar potentials, which is to support nonzero
chiral condensate in $v(z)$, even in free space with no chemical
potentials~\cite{KS07}. Therefore, we don't consider the scalar
sector in this work.

\subsection{Strange Matter}
We first study the strangeness-chemical-potential dependence of
hadronic observables. The mass of the kaon is given by
 \ba
m_{K^\pm} = \mp \tilde\mu_s + \sqrt{\tilde\mu_s^2 + m_0^2}, \ea
where $m_0$ is the mass with no strangeness chemical potential. When
$\tilde \mu_s$  increases, the masses of particles that contains
$\bar s$ quarks, $K^+$ and $K^0$, decrease while the masses of $K^-$
and $\bar K^0$, which have $s$ quarks, increase with $\tilde\mu_s$.

{}Similarly, we study the vector meson case and find that
 the masses of $K^{*-}$ and $\bar K^{*0}$ increase
and that those of $K^{*+}$ and $K^{*0}$ decrease with increasing
$\tilde \mu_s$: \ba m_{K^{*\pm}} = \mp \tilde\mu_s + m_0\, . \ea

Now, we study a transition from nuclear matter to strange matter in
a simple way. The boundary action, which is nothing but the 4D grand
canonical potential~\cite{DH}, is
 \ba
 S_{\rm b}&=&\frac{1}{2}M_5{\rm Tr} \int d^4x
\left(\frac{1}{z}V_0 \partial_z V_0\right)_{z=\epsilon}\,\no &=&
\frac{1}{2} M_5 \int d^4x \left(\frac{1}{z}(4\tilde{V}_0 \partial_z
\tilde{V}_0+2{\bar V}_0 \partial_z {\bar V}_0+V_0^3 \partial_z
V_0^3)\right)_{z=\epsilon}\,\no &=&M_5 \left (
4\tilde\mu_qc_q+2\tilde\mu_s c_s +\tilde\mu_Ic_I  \right ) \int
d^4x\, ,\label{4DS}
 \ea
where $c_i=12\pi^2 \rho_i/N_c$ with $i=q,s,I$, and $\rho_i$ is a
conjugate number density of $\tilde\mu_i$. Here, we turn on the
isospin chemical potential and study the transition from symmetric
nuclear matter to strange matter. To this end, we introduce \ba
&&\rho_Q  = 2\rho_q + \rho_s, \no
&&\rho_q=\frac{1}{2}(1-x)\rho_Q,\quad \rho_s = x\rho_Q \, . \ea As
previously mentioned, we cannot determine the relation between a
chemical potential and the corresponding number density in a
self-consistent way. Therefore, we have to borrow the relation from
other studies using lattice QCD or models of QCD. In this work, we
simply adopt the relation from a free quark gas: \ba \tilde\mu_q =
k_F^q = (\pi^2 \rho_q)^{1/3},\quad \tilde\mu_s = \sqrt{m_s^2 +
k_F^{s2}}=\sqrt{m_s^2 + (\pi^2 \rho_s)^{2/3}} \, . \ea The relevant
part of Eq. (\ref{4DS}) to study the transition is \bneqnas f(x)&=&
2\tilde\mu_q \rho_q +\tilde\mu_s \rho_s \no &=& (1-x) \rho_Q (\pi^2
\frac{1}{2} (1-x)\rho_Q)^{1/3}+x\rho_Q (m_s^2 + (\pi^2 x
\rho_Q)^{2/3})^{1/2}\, .
 \edeqnas
If $f(x)$ has an absolute minimum at a nonzero and positive $x$,
then we have a transition at that $x$. We calculate the transition
density at different strange quark masses to see how the transition
density changes with $m_s$.

The value of the fixed strange quark mass shows some discrepancy in
AdS/QCD studies~\cite{RP06, SW06, WX07, KS07}. The discrepancy is
mainly due to differences in the choices for the values of the
parameters and the backgrounds in the studies. For instance, one
finds $m_s=90~{\rm MeV}$ and $\la\bar ss\ra= (457.53~{\rm MeV})^3$
in the pure AdS~\cite{SW06} model, $m_s=138.5~{\rm MeV}$ and
$\la\bar ss\ra= (176~{\rm MeV})^3$ in a deformed AdS ~\cite{WX07},
and $m_s=40~{\rm MeV}$ and $\la\bar ss\ra= (333~{\rm MeV})^3$ in
pure AdS background ~\cite{KS07}.
 Note that in Ref.\cite{KS07}, $\la\bar uu\ra=\la\bar dd\ra=\la\bar ss\ra$ is
 assumed, so we choose values of $m_s$ in the range of 50 - 150 MeV, which are
introduced in other AdS Models. Then, the transition densities are
\bneqna \label{rC} && m_s = 50 ~{\rm MeV}, \quad \tilde
c_q^c=0.004\, , \no
&& m_s = 100 ~{\rm MeV}, \quad \tilde c_q^c=0.035 \, , \nonumber \\
&& m_s = 150 ~{\rm MeV}, \quad \tilde c_q^c=0.116\, . \edeqna Here,
we have introduced the dimensionless quark number density $\tilde
c_q = c_q z_m^3$. As expected, the transition density becomes larger
with heavier strange quark mass. We remark, however, that the
results in Eq.~(\ref{rC}) should be taken as being very schematic
because free quark gas has been assumed.

\section{Deconfinement Transition in Dense Matter}
In this section, we generalize the analysis done in
Ref.\cite{Herzog, KLNPS} to a system with various chemical
potentials. We note here that, unlike isospin matter, we turn on the
baryon chemical potential, as well as the others. Therefore, even
with no pion condensation, we can have $\rho_I\neq 0$.

We first summarize  \cite{Herzog, KLNPS}. The Euclidean
gravitational action is given by
\begin{equation} \label{action1}
S_{grav} ~=~ -\frac{1}{2\kappa^2} \int d^5x
\sqrt{g}\left(\textrm{R}+12\right),
\end{equation}
where $\kappa^2 $ is the 5D Newton constant.
 The cut-off thermal AdS (tAdS) is given by
\begin{equation}
ds^2=\frac{1}{z^2}\left(d\tau^2+dz^2+d\vec{x}^2_3\right),
\end{equation}
and the cut-off AdS black hole (AdSBH) reads
\begin{equation}
ds^2=\frac{1}{z^2}\left(f(z)d\tau^2+\frac{dz^2}{f(z)}+d\vec{x}^2_3\right),
\end{equation}
where $f(z)=1-(z/z_h)^4$, and $z_h$ is the horizon of the black
hole. The Hawking temperature is defined by $T=1/(\pi z_h)$. The
periodicity of the compactified Euclidean time direction of the
tAdS is given by
\begin{equation}
\beta^\prime = \pi z_h \sqrt{f(\epsilon)}.
\end{equation}

{}Now, we calculate the action density $V$, which is defined by the
action divided by the volume $d\vec{x}^2_3$. For the tAdS, we obtain
\begin{equation}
V_1 = \frac{4}{\kappa^2} \int^{\beta '}_{0} d\tau
\int^{z_{m}}_{z_0}\frac{dz}{z^5}\, ,
\end{equation}
and for the AdSBH, we have
\begin{equation}
V_2 = \frac{4}{\kappa^2}\int^{\pi z_h}_{0} d\tau
\int^{min(z_{m},z_h)}_{z_0}\frac{dz}{z^5}\, .
\end{equation}
Then, we arrive at
\begin{equation}
\Delta V_g = V_2 -V_1= \left\{\begin{array}{ll} \frac{ \pi
z_h}{\kappa^2}
\frac{1}{2z_h^4} & z_{m} < z_h\\ \\
\frac{ \pi z_h}{\kappa^2} \left( \frac{1}{z_{m}^4} -
\frac{1}{2z_h^4}\right) & z_{m} > z_h.\end{array}\label{dVg} \right.
\end{equation}
 When $\Delta V_g$ is positive (negative), the thermal AdS (the AdS black hole) is stable~\cite{Herzog}.
Thus, at $\Delta V_g=0$, there exists a Hawking-Page transition. In
the second case, $z_{m} > z_h$, the Hawking-Page transition occurs
at
\begin{equation}    \label{tempads}
T_0 = 2^{1/4}/(\pi z_{m})\, .
\end{equation}
This is the result in Ref.\cite{Herzog} for the hard wall model.

To study the Hawking-Page transition at finite density~\cite{KLNPS},
we consider the meson action given in~Eq. (\ref{S5}). The Euclidean
action for mesons is given by
\begin{equation} \label{action2}
S_{matter} ~=~  M_5 \int d^5x \sqrt{g}~ \textrm{Tr}
\left[\frac{1}{2}|D_\mu \Phi|^2 + \frac{1}{2} M_{\Phi}^2|\Phi^2|
+\frac{1}{4}\left(L_{MN}L^{MN}+R_{MN}R^{MN}\right)\right]\, .
\end{equation}
The solution of the equation of motion for the (Euclidean) time
component of the U(1) bulk vector field is given by \ba
V_\tau=c_1+c_2z^2\, . \ea In Ref.\cite{KLNPS}, $c_1$ is identified
as the baryon (quark) chemical potential, and $c_2$ is nothing but
the baryon number density. {}Following the same procedure as for the
gravity action, we obtain the regularized action density as follows:
\begin{equation}
V_{v1} = \pi z_h M_5 N_f  c_2^2 z_{m}^2
\end{equation}
for the tAdS and
\begin{equation}
V_{v2}=\left \{\begin{array}{ll} \pi z_h M_5 N_f c_2^2 z_{h}^2 & z_h<z_{m}\\
\\\pi z_h M_5 N_f c_2^2 z_{m}^2 & z_h>z_{m} \end{array} \right.
\end{equation}
for the AdSBH. Then, the difference of the actions reads
\begin{equation}
\Delta V_{v}=V_{v2}-V_{v1}=\left\{\begin{array}{ll} -\pi z_h M_5 N_f
c_2^2(z^2_{m}- z_{h}^2) & z_h<z_{m}\\ \\0 & z_h>z_{m}. \end{array}
\right. \label{dVv}
\end{equation}
Combining Eq. (\ref{dVg}) and Eq. (\ref{dVv}), we
obtain~\cite{KLNPS},  for $z_h<z_{m}$,
\begin{equation}
\Delta V = \frac{ \pi z_h}{\kappa^2}\left[ \frac{1}{z_{m}^4}
-\frac{1}{2z_h^4} -\frac{ N_f c^2_2}{48N_c}\left(z^2_{m}-
z_{h}^2\right)\right]\, ,\label{dVt}
\end{equation}
where we have used
\begin{equation}
\frac{1}{\kappa^2}=\frac{1}{8\pi G_5}, ~~~~
\frac{1}{G_5}=\frac{32N^2_c}{\pi L^3}, ~~ ~~
\textrm{and}~~~~M_5=\frac{N_c}{12\pi^2 L}\, .
\end{equation}
Here, the second relation is taken from Ref.\cite{Csaki:2006ji}.
{}From Eq.~(\ref{dVt}), it is obvious that $T_c\equiv 1/(\pi
z_h^0)$, where $z_h^0$ is a solution of $\Delta V=0$,
 will depend on $c_2$ or the baryon number density.

Now, we extend the analysis done in Ref.\cite{KLNPS} with the
isospin and the strangeness chemical potentials. For this, we
generalize Eq. (\ref{dVt}) to obtain
\begin{equation}
\Delta V = \frac{ \pi z_h}{\kappa^2}\left[ \frac{1}{z_{m}^4}
-\frac{1}{2z_h^4} -\frac{ 1}{48N_c}\left( 4c_q^2+2c_s^2+c_I^2
\right)\left(z^2_{m}- z_{h}^2\right)\right]\, .\label{dVF}
\end{equation}

We consider a few situations that may be relevant to the physics of
relativistic heavy-ion collisions and neutron stars (or quark
stars~\cite{Itoh:1970uw}).

First we consider the case where  $c_I=0$ with $c_q$ and $c_s$ being
nonzero. Here, we introduce $c_b=2c_q + c_s$ and $c_s={\rm
constant}$. We assume that the weak interaction time scale is long
enough to be ignored and that the strangeness is conserved. This
situation could correspond to the initial stages of a heavy-ion
collision, where the initially produced strangeness is conserved.
\begin{figure}[!ht]
\begin{center}
{\includegraphics[angle=0, width=0.7\textwidth]{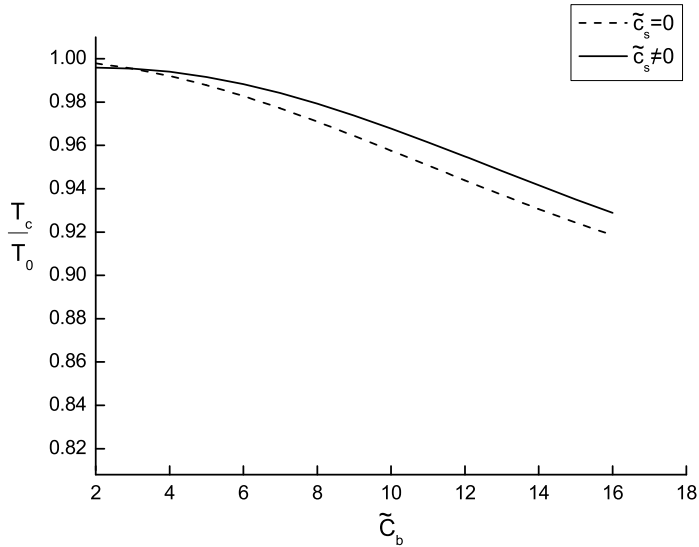}} \caption{
\label{Tcst} \small The critical temperature at finite strange
density. The solid line is for $\tilde c_s\neq 0$ while the dashed
line is for $\tilde c_s=0$, where $\tilde c_s = c_s z_m^3$ and
$\tilde c_b = c_b z_m^3$.}
\end{center}
\end{figure}
As can be seen in Fig.~\ref{Tcst}, we observe that the deconfinement
temperature with finite strangeness density is higher than it is
with zero strangeness density.

\begin{figure}[!ht]
\begin{center}
{\includegraphics[angle=0, width=0.7\textwidth]{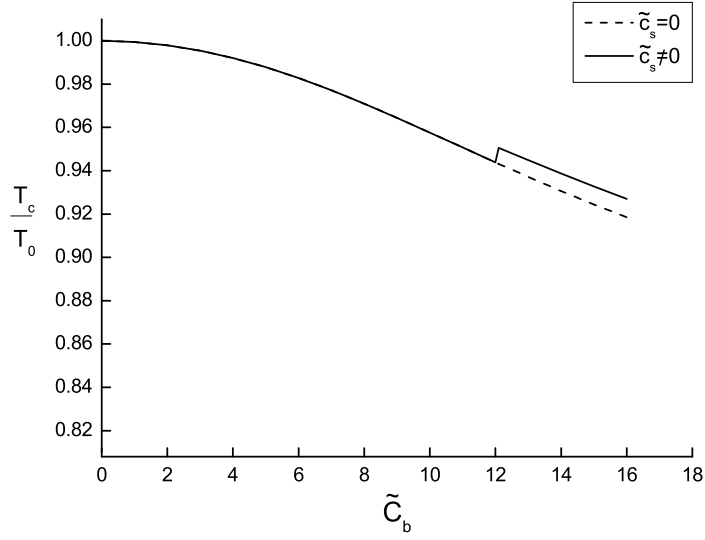}}
\caption{
 \label{Tcs} \small The critical temperature at the phase transition
to strange matter. The solid line is for the case where the phase
transition occurs at $ \tilde c_h $ while the dashed line is for the
case with no transition. We assume $\tilde c_h=12$. }
\end{center}
\end{figure}

Second, we consider the case where  $c_I=0$ with $c_q$ and $c_s$
being nonzero and where the weak processes and charge conservation
have to be considered. In studies based on QCD effective
theories~\cite{PHZSG}, the transition to strange matter takes place
at a density around $2-3$ times that of normal nuclear matter, and
the transition is first order.   To simulate such conditions, we
introduce $c_s=yc_b$, where $0\le y\le 1$. We further assume that
the transition occurs at some density $c_h$ and that at this
density, $y$ acquires a finite value. For example, we take $y=0.1$,
which is close to the value in Ref.\cite{PHZSG}, to obtain
Fig.~\ref{Tcs}.
An interesting feature observed in Fig.~\ref{Tcs} is that the
first-order nature of the transition to strange matter leads to a
cusp in $T_c$.

\begin{figure}[!ht]
\begin{center}
{\includegraphics[angle=0, width=0.7\textwidth]{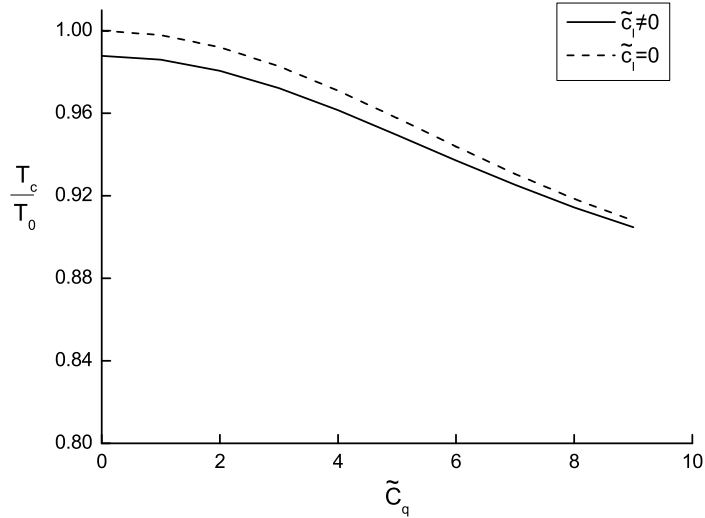}} \caption{
 \label{Tci} \small The critical temperature at finite isospin. The
dashed line is for $\tilde c_I=0$ while the solid line is for
$\tilde c_I\neq 0 \left(\tilde c_I = c_I z_m^3\right)$.}
\end{center}
\end{figure}

Finally, we consider the case where $c_s=0$  and  $c_I\neq0$. In
this case, the critical temperature decreases as compared to the
case with $c_I=0$, as shown in Fig.~\ref{Tci}. Similar results are
observed in a model calculation based on a hadron resonance
gas~\cite{Toublan:2004ks}.

Closing this section, we discuss possible corrections, other than
the one studied in this work, to the Hawking-Page transition
temperature $T_0$. Note that $T_0$ is the transition temperature
determined with the gravity action only. As is well-known, the
matter action is $1/N_c$ sub-leading compared to the leading gravity
action, so the correction to $T_0$ from the matter action is
suppressed by $1/N_c$, as shown in Eq.~(\ref{dVF}). There are,
however, other sources of the $1/N_c$ corrections to $T_0$. First of
all, we need to consider the effect of the back-reaction due to the
matter, such as the chemical potentials and corresponding densities
on the background. So far, however, a full back-reacted background
in confined phase due to the chemical potentials and corresponding
densities has not been found. We refer to Ref.\cite{Deformed} for
back-reacted backgrounds in some other situations. Another source of
a correction could be from 5D loops, which are completely ignored in
the present work. This defect, however, seems common to studies
based on AdS/CFT or AdS/QCD. Although we may not be able to
calculate the corrections from the back-reaction and 5D loops in a
simple manner, we have to include them to obtain complete
corrections to $T_0$ due to the matters. Therefore, we may conclude
that in the present work, we have calculated some part of the
leading $1/N_c$ corrections due to the matters.

\section{Summary}
Using the AdS/QCD approach, we studied isospin and strange matters,
their effects on the hadronic observables, and the critical
temperature for the deconfinement transition. The isospin matter can
serve as a good testing ground for the AdS/QCD approach because it
has been well-studied by using lattice QCD, by using chiral
perturbation theory, and by using effective models of QCD, including
the NJL model. Therefore, we first considered the isospin matter and
found that the mass of $\pi^+$ increases with $|\tilde\mu_I|$ while
that of $\pi^-$ decreases, which is in agreement with the results
from lattice QCD and from other studies at  small isospin chemical
potential. When the isospin chemical potential is high enough, one
may expect the pion condensation observed in the previous
studies~\cite{SSiso, NJL, Lattice, Other}. In the chiral limit,
$m_0=0$, we showed that the pion condensation occurs in the present
study, which is consistent with the observations made in various
studies~\cite{SSiso, NJL, Lattice, Other} and in a top-down
approach~\cite{APSZ}. With nonzero $m_0$, however, the mass of
$\pi^-$ obtained in the present work is always non-zero unless
$\tilde\mu_I$ goes to infinity. Therefore, pion condensation can
exist only if $\tilde\mu_I=\infty$, which is not the case in
Ref.\cite{SSiso, NJL, Lattice, Other}. We may attribute this
discrepancy to the fact that we don't take into account the
back-reaction due to the isospin chemical potential on the metric,
especially when the chemical potential is large. We observed that
the space component of the pion decay constant and the masses of
$\rho^-$ and $a_1$  decrease with increasing $\tilde\mu_I$ while the
$\rho$ and $a_1$ decay constants are independent of $\tilde\mu_I$.

Then, we calculated the $\tilde\mu_s$ dependence of the hadronic
observables and found that the dependence is similar to that of
$\tilde\mu_I$. We also investigated the transition from nuclear
matter to strange matter in our model. The relation between the
chemical potential and the density, however, could not be obtained
within the present approach. For simplicity, we adopted the relation
from a free quark gas and estimated the critical density for the
transition. We found that the critical density was very sensitive to
the mass of the strange quark, as expected.

Finally, we estimated the critical temperature of the deconfinement
transition  for different isospin  and strangeness number densities.
For the strangeness, we considered two cases that might be relevant
to relativistic heavy-ion collisions and neutron stars,
 which are summarized in Figs. \ref{Tcst}  and \ref{Tcs}.
At a fixed baryon number density with a finite isospin chemical potential,
we found that $T_c$ in asymmetric nuclear matter, $\tilde c_I\neq 0$, is smaller than that
in symmetric nuclear matter, $\tilde c_I=0$.

\vskip 0.3cm

{\bf ACKNOWLEDGEMENTS}\\

 The work of SHL and KK was supported  by the Korea Research Foundation (KRF-2006-C00011).
YK acknowledges the Max Planck Society (MPG) and the Korea Ministry
of Education, Science, and Technology (MEST) for their support of
the Independent Junior Research Group at the Asia Pacific Center for
Theoretical Physics (APCTP).

\end{document}